\documentclass[runningheads]{llncs}
\usepackage[utf8]{inputenc} %
\usepackage[T1]{fontenc}    %
\usepackage{hyperref}       %
\usepackage{url}            %
\usepackage{booktabs}       %
\usepackage{amsfonts}       %
\usepackage{nicefrac}       %
\usepackage{microtype}      %
\usepackage[table]{xcolor}         %
\usepackage{graphicx}
\usepackage{subcaption}
\usepackage{amsmath}
\usepackage{makecell}
\usepackage{adjustbox}

\begin{document}
\title{Physics-Informed Latent Diffusion for Multimodal Brain MRI Synthesis}
\author{Sven Lüpke$^*$ \inst{1} \and
Yousef Yeganeh$^*$ \inst{1,2} \and
Ehsan Adeli \inst{3} \and
Nassir Navab \inst{1,2} \and
Azade Farshad \inst{1,2}
}

\authorrunning{S. Lüpke et al.}
\def\thefootnote{*}\footnotetext{Equal contribution}
\institute{Technical University of Munich \and
Munich Center for Machine Learning \and
Stanford University}

\maketitle              %
\begin{abstract}
Recent advances in generative models for medical imaging have shown promise in representing multiple modalities. However, the variability in modality availability across datasets limits the general applicability of the synthetic data they produce. To address this, we present a novel physics-informed generative model capable of synthesizing a variable number of brain MRI modalities, including those not present in the original dataset. Our approach utilizes latent diffusion models and a two-step generative process: first, unobserved physical tissue property maps are synthesized using a latent diffusion model, and then these maps are combined with a physical signal model to generate the final MRI scan. Our experiments demonstrate the efficacy of this approach in generating unseen MR contrasts and preserving physical plausibility. Furthermore, we validate the distributions of generated tissue properties by comparing them to those measured in real brain tissue.

\end{abstract}

\section{Introduction}

Synthetic data generation has emerged as a critical research area in the medical domain, where real data samples are often scarce due to limited availability, strict privacy regulations, or ethical considerations. State-of-the-art (SOTA) generative models for medical images increasingly employ denoising diffusion models \cite{sohl2015deep,ho2020denoising,pinaya2022brain}, which generate images by iteratively denoising Gaussian noise. Due to the high computational demands of these models, the latent diffusion model (LDM) \cite{rombach2022high} performs denoising in a lower dimensional latent space by first encoding images with a variational autoencoder (VAE). 
The increasing integration of multimodal data \cite{hatamizadeh2021swin} necessitates the creation of multimodal synthetic datasets. Recent multimodal generative models in the medical imaging domain are limited to the fixed set of modalities used to train these models \cite{zhou2021feature,na2023generation}. However, the availability of modalities varies drastically across different datasets in the medical domain, limiting the general applicability of the data synthesized by these models.

Multimodal variational autoencoders \cite{suzuki2022survey} can process a varying number of modalities by aggregating unimodal inference distributions of the individual modalities into a joint multimodal distribution. Shi et al. \cite{shi2019variational} propose to model the inference distribution as a mixture of experts, creating shared and private sub-spaces in the joint distribution. Alternatively, the joint multimodal distribution can be modeled by a product of unimodal experts \cite{hinton2002training}. Joshi et al. \cite{joshi2022generalized} generalized the product of experts for multimodal inference in the presence of noise by adaptively weighting each modality. 

Physics-informed methods \cite{raissi2019physics} combine neural networks with physical equations, ensuring their predictions respect the given physical laws. In contrast to purely data-driven solutions, they offer improved extrapolation capabilities \cite{davini2021using}. Trask et al.~\cite{trask2022unsupervised} integrate a product-of-experts into a physics-informed framework for disentangling multimodal data in the latent space using a Gaussian mixture prior over the latent variables.

In this paper, we propose a novel physics-informed diffusion-based generative model for multimodal brain MR scans. Inspired by quantitative MRI techniques \cite{jara2022primary,jacobs2023generalizable}, we utilize the acquisition parameters \cite{denck2021mr} in combination with a physical signal model \cite{borges2023unsupervised} and a latent diffusion model to synthesize images in a two-step process: (1) The generation of modality shared physical tissue property maps, namely the proton density, the longitudinal relaxation time, and the transverse relaxation time, (2) The application of a physical signal model with a desired set of scanner acquisition parameters to the tissue property maps to obtain the final MRI scan.

We showcase the versatility of our approach by generating MRI contrasts not present in the training data. This is achieved by combining generated tissue property maps with acquisition parameters unseen during training, effectively expanding the variety of images while preserving physical plausibility. Additionally, we compare the distributions of generated tissue properties against those measured in real brain tissue, highlighting the potential for future research on a unified generative model for multimodal medical images.

\section{Method}
\subsection{MRI Signal Model}

The signal intensity in an MRI scan is mainly influenced by two factors: the scanner configuration and the tissue properties. Given the tissue properties, an MRI simulator can generate a variety of MR contrasts based on the chosen MR sequence and the acquisition parameters. However, due to the high computational cost of a full MRI simulation, the relationship between tissue properties, which are given by the proton density (PD), the longitudinal relaxation time (T1) and the transverse relaxation time (T2), and the signal intensity is instead modeled by a set of signal equations. These equations only depend on a subset of all possible acquisition parameters, namely the echo time (TE), the repetition time (TR), and the inversion time (TI). For T1-weighted MPRAGE \cite{deichmann2000optimization}, T2-weighted spin-echo (SE) \cite{hashemi2012mri} and FLAIR \cite{hashemi2012mri} sequences, the equations modeling the signal intensity $s$ at a spatial location $x$ are given by:

\begin{equation}
\begin{split}
    s_{MPRAGE}(x) &= G \cdot PD(x) \bigg(1 - \frac{2e^{\frac{-TI}{T1(x)}}}{1 + e^{\frac{-TR}{T1(x)}}} \bigg) \\
    s_{SE}(x) &= G \cdot PD(x) \bigg(1 - e^{\frac{-TR}{T1(x)}}\bigg)e^{\frac{-TE}{T2(x)}} \\
    s_{FLAIR}(x) &= G \cdot PD(x) \bigg(1 - 2e^{\frac{-TI}{T1(x)}} + e^{\frac{-TR}{T1(x)}}\bigg)e^{\frac{-TE}{T2(x)}}
    \label{eq:signal}
\end{split}
\end{equation}

The tissue properties T1, T2, and PD depend on the location $x$, whereas $G$ is a global parameter that models the scanner gain. Our model ignores the scanner gain and assumes $G = 1$. Since the signal depends linearly on PD and G, any scanner-specific signal scaling is absorbed into the proton density map.

\begin{figure}[t]
    \centering
    \includegraphics[width=\textwidth]{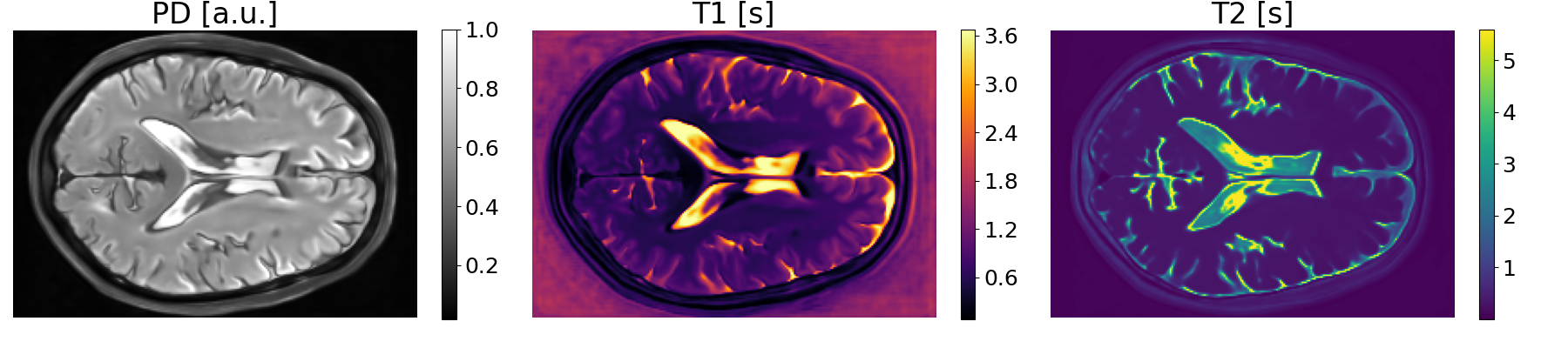}
    \caption{Unobserved tissue property maps generated by our model.}
    \label{fig:combined}
\end{figure}

\subsection{Multimodal Physics-Informed Variational Autoencoder}

To efficiently perform denoising diffusion to generate tissue property maps containing PD, T1, and T2, we design a multimodal physics-informed variational autoencoder that downsamples $N$ multimodal MRI scans by a factor of 8 into a shared lower dimensional latent representation.

We use a single shared convolutional encoder to encode the input modalities independently into unimodal latent distributions $q_{\phi}(z|x_i)$. The encoder is conditioned on the acquisition parameters corresponding to each input image using adaptive group normalization \cite{dhariwal2021diffusion} in the residual blocks:
\begin{equation}
    AdaGN(h, s, b) = s \cdot GroupNorm(h) + b
\end{equation}
where $h$ is the hidden feature and $(s, b) \in \mathbb{R}^{2 \times N} = MLP(TE, TR, TI)$ is the output of a multilayer perceptron given the acquisition parameters.

To combine the unimodal latent distributions $q_{\phi}(z|x_i)$ into a shared multimodal distribution $q(z|X)$, we use a product-of-experts:
\begin{equation}
\label{eq:prod_ex}
    q(z|X) \propto \prod_{i=1}^N q_{\phi}(z|x_i)
\end{equation}
The unimodal distributions $q_{\phi}(z|x_i)$ are assumed to be Gaussian, with their mean $\mu_i$ and standard deviation $\sigma_i$ being predicted by the encoder. Thus, the product of the unimodal distributions is also a Gaussian and is given by:

\begin{equation}
    q(z|X) = \mathcal{N}(\mu, \sigma^2)
\end{equation}
\begin{equation}
    \sigma^{-2} = \sum_{i=1}^{N} \sigma_i^{-2}, \hspace{0.5cm} \mu = \sigma^2 \sum_{i=1}^{N} \frac{\mu_i}{\sigma_i^{2}}
\end{equation}

\begin{figure}[t]
    \centering
    \includegraphics[width=\textwidth]{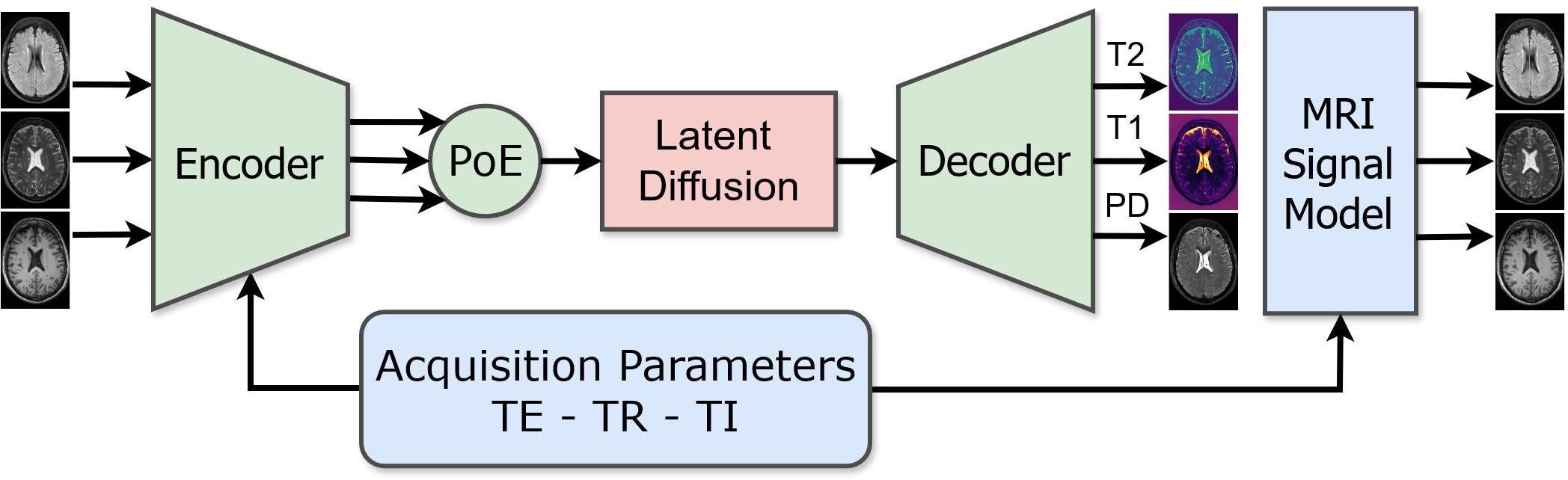}
    \caption{Overview of our physics-informed generative model. We combine an MRI signal model with a product-of-experts (PoE) multimodal variational autoencoder and a latent diffusion model.}
    \label{fig:arch}
\end{figure}

We decode the shared latent representation using a convolutional network into tissue property maps by passing its output through the exponential function, ensuring that the property values are greater than zero, and transforming the resulting values by the signal model (\autoref{eq:signal}) to reconstruct the input.
Following previous work \cite{rombach2022high,pinaya2022brain}, we combine an L2 reconstruction loss with a perceptual loss, a patch-wise adversarial loss, and a KL-regularization loss on the latent distribution. 

\begin{figure}[t]
    \centering
    \begin{subfigure}{\linewidth}
        \caption{MPRAGE}
        \includegraphics[width=\textwidth]{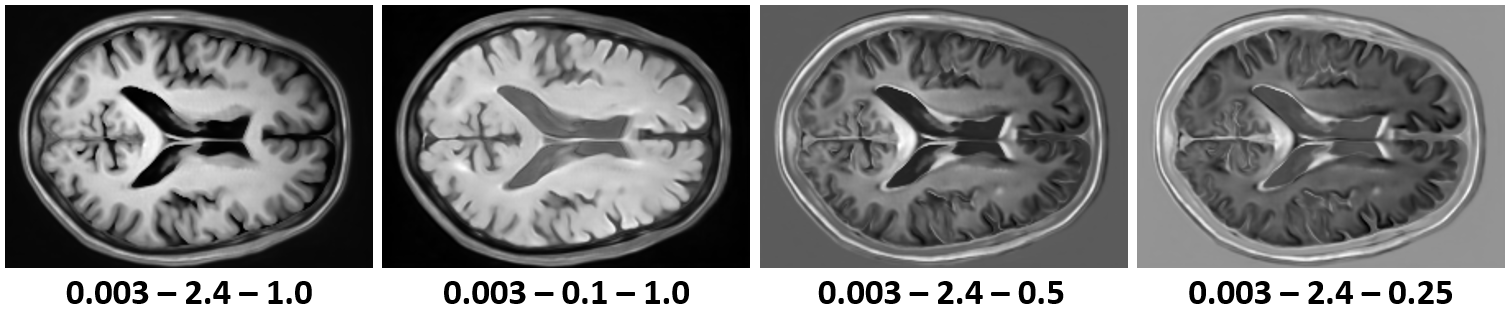}
    \end{subfigure}
    \hfill
    \begin{subfigure}{\linewidth}
        \caption{Spin Echo}
        \includegraphics[width=\textwidth]{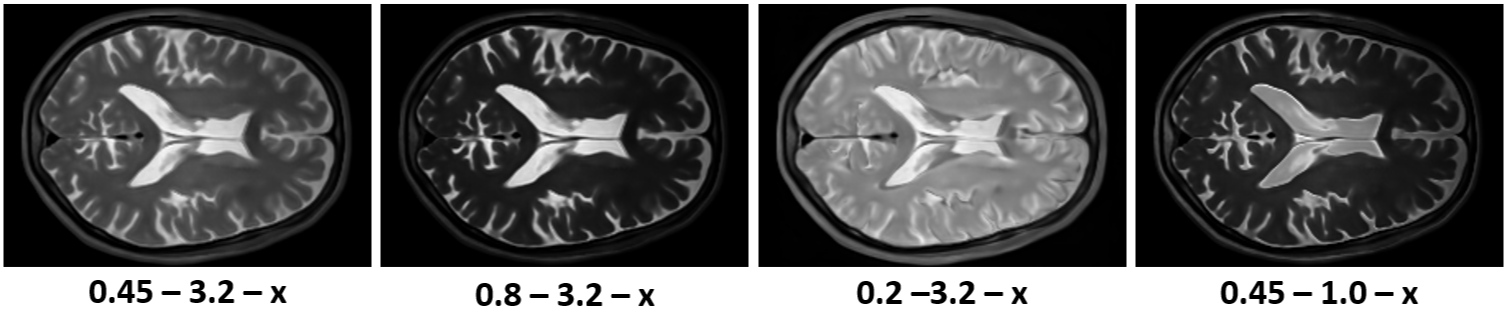}
    \end{subfigure}
    \hfill
    \begin{subfigure}{\linewidth}
        \caption{FLAIR}
        \includegraphics[width=\textwidth]{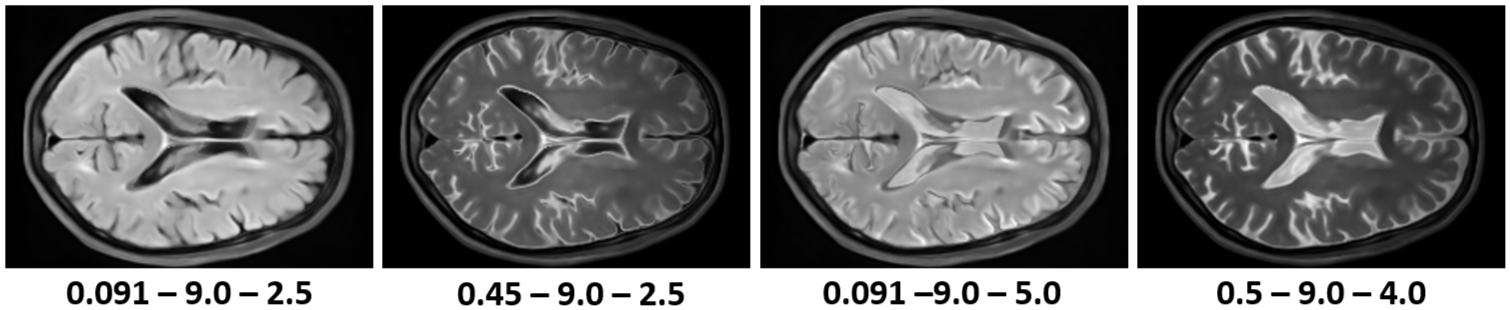}
    \end{subfigure}
    \caption{Images of a single brain, represented by the tissue property maps in \autoref{fig:combined} generated with different signal models and varying acquisition parameters TE-TR-TI. Spin-Echo sequences do not use inversion recovery. The leftmost images use parameters commonly found in the OASIS-3 dataset, whereas the other images use parameter combinations not present in the training data.}
    \label{fig:generated_parameter_ablation}
\end{figure}

Because the tissue property values that reconstruct the input are not unique, we employ a prior over T1 and T2. More specifically, we employ a light L2-regularization on the output values $o_{T1}$ and $o_{T2}$ of the decoder and apply a constant bias before passing them through the exponential function:
\begin{align}
    T1 &= exp(o_{T1} + b_{T1}) \\
    T2 &= exp(o_{T2} + b_{T2})
    \label{eq:exp}
\end{align}
We set $s_{T1} = ln(1)$ and $s_{T2} = ln(0.1)$ based on the median of the property value distribution measured in real brain tissue \cite{bojorquez2017normal}, corresponding to a log-normal prior with a median of $1$ on T1 and a median of $0.1$ on T2.

\begin{figure}[t]
    \centering
    \begin{subfigure}[b]{0.45\textwidth}
        \centering
        \includegraphics[width=\textwidth]{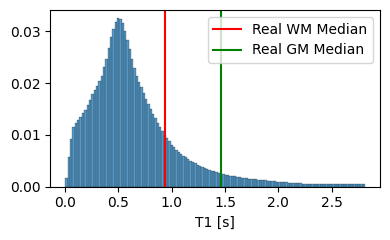}
        \label{fig:image1}
    \end{subfigure}
    \hfill
    \begin{subfigure}[b]{0.45\textwidth}
        \centering
        \includegraphics[width=\textwidth]{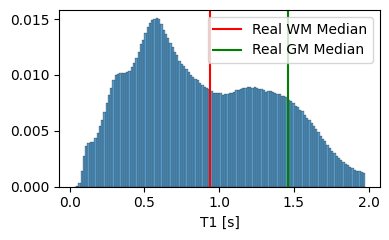}
        \label{fig:image2}
    \end{subfigure}
    \begin{subfigure}[b]{0.45\textwidth}
        \centering
        \includegraphics[width=\textwidth]{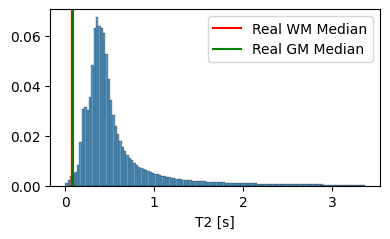}
        \caption{Without Prior}
        \label{fig:image3}
    \end{subfigure}
    \hfill
    \begin{subfigure}[b]{0.45\textwidth}
        \centering
        \includegraphics[width=\textwidth]{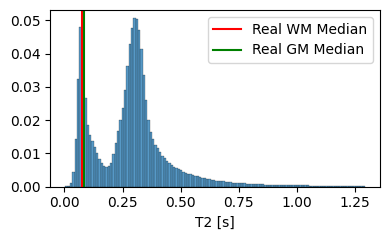}
        \caption{With Prior}
        \label{fig:image4}
    \end{subfigure}
    
    \caption{Distribution of T1 and T2 values in the generated images compared to the median T1 and T2 of white matter (WM) and grey matter (GM) in real brains reported by \cite{bojorquez2017normal}. For visualization purposes, we clipped the distributions of the generated properties to the 95\textsuperscript{th} percentile. The results show that the introduction of the prior leads to the generation of more realistic property distributions.}
    \label{fig:main}
\end{figure}

\subsection{Latent Diffusion Model}

To generate new tissue properties maps, we train a latent diffusion model \cite{rombach2022high} on the joint multimodal representation we obtain through \autoref{eq:prod_ex}. Given a sample $z \sim q(z|X)$, the objective function is:
\begin{equation}
    L_{LDM} = \mathbb{E}_{z, \epsilon \sim \mathcal{N}(0, 1), t} \Big[ ||\epsilon - \epsilon_{\theta}(z_t, t)||_2^2 \Big]
\end{equation}
where $\epsilon_{\theta}$ is a UNet that predicts the noise added at each timestep $t$.

\section{Experiments and Results}

\subsubsection{Dataset and Preprocessing}

Our experiments used T1-weighted MPRAGE, T2-weighted spin-echo, and FLAIR scans from the OASIS-3 dataset \cite{lamontagne2019oasis}. The MRI volumes were co-registered to a common MNI space using UniRes \cite{brudfors2019tool} and cropped to a resolution of $160 \times 224 \times 160$. Due to memory limitations, we did not use the entire brain volumes but only the 2D axial slices, each having a resolution of $224 \times 160$. At the input of the VAE, we scaled the image value to the range [-1, 1], only considering the 99.5\textsuperscript{th} percentile of the pixel intensities due to the long upside tails in the signal intensity distribution of MR images. As targets for reconstruction during the VAE training, we used unscaled images because our signal model is designed to predict the raw signal intensity produced by the scanner.

\subsubsection{Training Details}

\begin{table}[t]
\centering
\begin{adjustbox}{max width=\textwidth}
\rowcolors{2}{gray!25}{white}
\begin{tabular}{cccc|cccc}
     \textbf{G Regression} & \makecell{\textbf{Modality} \\ \textbf{Dropout}} & \makecell{\textbf{T1 \& T2} \\ \textbf{Prior}} & \textbf{AdaGN} & \textbf{MSE $\downarrow$} & \textbf{MAE $\downarrow$} & \textbf{MS-SSIM $\uparrow$} &  \textbf{PSNR $\uparrow$}\\
     \hline
     \checkmark & \checkmark & \checkmark & \checkmark & 1542 & 25.35 & 0.9577 & 30.89 \\
     & \checkmark & \checkmark & \checkmark & 1268 & 22.94 & 0.9603 & 31.67 \\
     \checkmark & & \checkmark & \checkmark & 1367 & 24.11 & 0.9609 & 31.39 \\
     & & \checkmark & \checkmark & \textbf{996} & \textbf{20.25} & \textbf{0.9632} & \textbf{32.70} \\
     \checkmark & \checkmark &  & \checkmark & 1592 & 25.67 & 0.9565 & 30.75 \\
      & \checkmark &  & \checkmark & 1365 & 23.63 & 0.9528 & 31.38 \\
     \checkmark & & & \checkmark & 1401 & 24.12 & 0.9607 & 31.30 \\
     & & & \checkmark & 1084 & \underline{20.76} & \underline{0.9617} & \underline{32.40} \\
     \checkmark & \checkmark & \checkmark & & 1531 & 25.42 & 0.9554 & 30.91 \\
     & \checkmark & \checkmark & & 1179 & 21.67 & 0.9596 & 32.00 \\
     \checkmark & & \checkmark & & 1436 & 24.64 & 0.9565 & 31.18 \\
     & & \checkmark & & \underline{1083} & 20.85 & 0.9613 & 32.35 \\
     \checkmark & \checkmark &  & & 1635 & 25.94 & 0.9527 & 30.63  \\
      & \checkmark &  & & 1391 & 23.58 & 0.9578 & 31.37 \\
     \checkmark & &  & & 1537 & 25.16 & 0.9579 & 30.90 \\
     & & & & 1388 & 23.65 & 0.9577 & 30.89 \\
\end{tabular}
\end{adjustbox}
\caption{Quantitative results of our ablation study, showing the reconstruction performance of our physics-informed VAE model. The best metrics are marked \textbf{bold}, and the second best is \underline{underlined}.}
\label{tab:metrics}
\end{table}

For training the networks, we mainly adopt the hyperparameters used by Pinaya et al. \cite{pinaya2022brain} and train all networks with the Adam optimizer. We trained the VAE for 50 epochs using a batch size of 4, where the encoder and decoder used a learning rate of $5e-5$ and the discriminator used a learning rate of $1e-4$. The latent diffusion UNet was trained with a learning rate of $2.5e-5$ for 100 epochs using a batch size of 16. All models were trained on a Nvidia Titan Xp GPU with 12 GB of VRAM.

\subsection{Results}

\subsubsection{Synthesis of Unseen Modalities}

A key advantage of our physics-informed generative model over previous methods is the ability to synthesize modalities not present in the training data.
\autoref{fig:generated_parameter_ablation} shows how changes in the acquisition parameters affect the generated images using different signal models given a single tissue property map generated by our model. Although we can generate various novel MR contrasts, some acquisition parameter combinations can still lead to unrealistic images with bright backgrounds.

\subsubsection{Analysis of the Tissue Property Distribution}

To ensure that our model can extrapolate to unseen modalities, it is crucial that the tissue property maps contain physically plausible values. In \autoref{fig:main}, we compare the distribution of T1 and T2 values generated by our model to the average T1 and T2 values of white matter and grey matter in actual brain tissue. The results highlight the importance of the prior on the predicted T1 and T2 values, as it guides the model towards generating tissue properties that are, on average, more realistic. However, it is essential to note that the generated distribution includes tissue types beyond white matter and grey matter, such as cerebrospinal fluid and bones.

\begin{figure}[t]
    \centering
    \begin{subfigure}[b]{\textwidth}
        \centering
        \includegraphics[width=\textwidth]{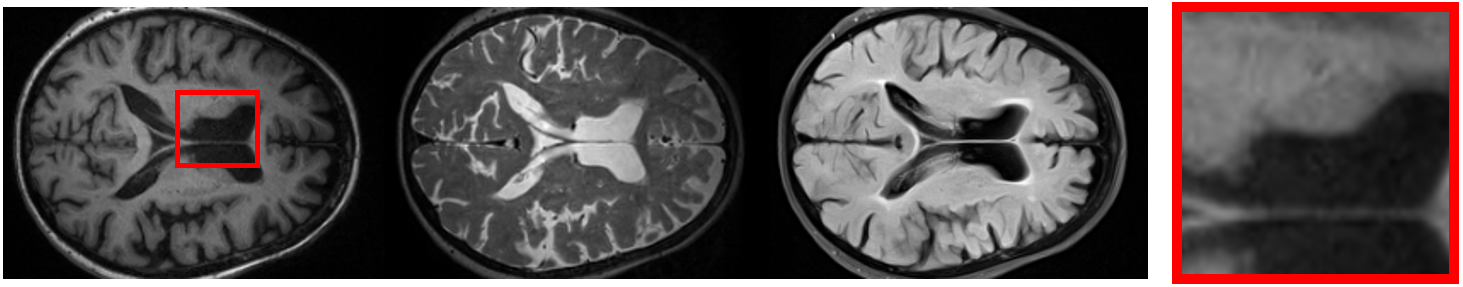}
    \end{subfigure}
    \hfill
    \begin{subfigure}[b]{\textwidth}
        \centering
        \includegraphics[width=\textwidth]{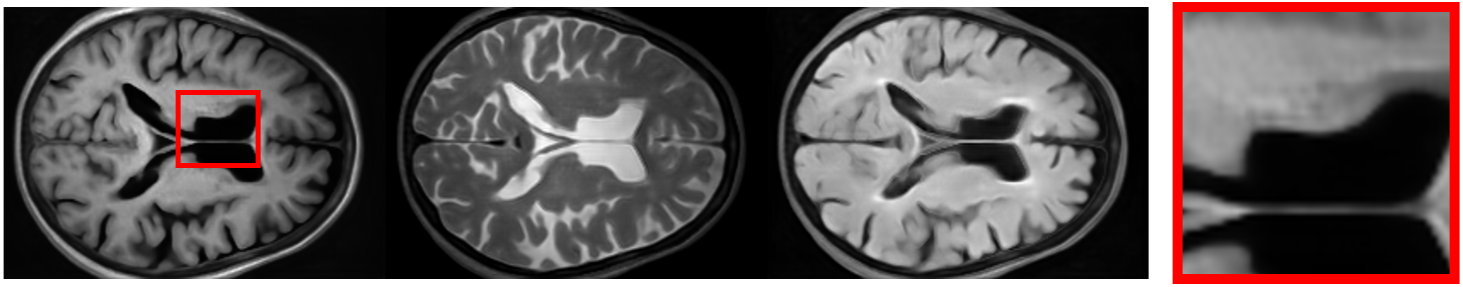}
    \end{subfigure}
    \caption{Images from the data set (top) and their physics-informed VAE reconstructions (bottom), showing the absence of scanner noise in the images generated by our model.}
    \label{fig:recon_noise}
\end{figure}

\subsubsection{Reconstruction Performance}

We evaluated the impact of the encoder conditioning via adaptive group normalization and the prior on T1 and T2 on the reconstruction performance of our VAE model measured by the mean squared error (MSE), the mean absolute error (MAE), the multi-scale structural similarity index (MS-SSIM), and the peak signal-to-noise ratio (PSNR). The results in \autoref{tab:metrics} show that encoder conditioning improves the model's performance in the majority, whereas the prior improves its performance in all cases.

We also experimented with modality dropout, a method commonly employed by multimodal models \cite{zhou2021feature,borges2023unsupervised}, which randomly removes modalities from the input to encourage the encoder to extract information from all modalities. However, this did not provide any benefits, which can likely be attributed to our use of the product-of-experts for multimodal fusion. 
Additionally, we attempted to regress the scanner gain $G$ using a convolutional neural network, which has not benefited our model's performance either.

\subsection{Limitations}

The employed signal model only considers a perfect MR scanner that does not add any noise to images. Thus, the MRI scans generated by our model do not contain any noise either \autoref{fig:recon_noise}. While this might be desirable for some applications, noise is required to create realistic MRI scans. One solution could be to model the residual noise separately using another diffusion model conditioned on the noise-free MRI scan \cite{mardani2024residual}. Additionally, the accuracy of the tissue property maps is limited by the number of available MR contrasts per scanning session. This issue could be addressed by introducing location-dependent priors based on the expected tissue type, such as white matter, grey matter, and cerebrospinal fluid. A more robust evaluation of the generated tissue properties would furthermore require access to real tissue property maps, which advanced quantitative MRI techniques can obtain \cite{jara2022primary}.

\section{Conclusion}

We have presented a physics-informed approach to the generative modeling of MRI scans. By combining physical MR signal models with a variational autoencoder and a latent diffusion model, we separated the generative process into two steps: the generation of unobserved physical tissue properties and a scanner model that transforms the tissue properties to signal intensities in the final MRI scans. Our results show that our model can generate MRI modalities beyond those seen by the model during training. However, the precise evaluation of the intermediate tissue property maps remains challenging. In the future, our approach could be improved with more sophisticated MRI signal models. It could theoretically be extended to develop physics-informed generative models for other medical imaging modalities beyond MRI, such as computed tomography or ultrasound scans.

\bibliographystyle{plain}
\bibliography{references}

\end{document}